\documentstyle[12pt,rotate,portland,psfig,epsf,aasms4]{article}
\input{rotate}
\received{}
\accepted{}
\journalid{}{}
\articleid{}{}

\renewcommand{\deg}{\mbox{$^{\circ}$}} 

\renewcommand{\aa}{{A\&A}}
\renewcommand{\apj}{{ApJ}} 
\newcommand{\Chandra}{\mbox{\it Chandra}}
\newcommand{\chandra}{\mbox{\it Chandra}}

\newcommand{\etal}{{\it et al.}} 

\newcommand{\iauc}{{IAU Circ.}}

\newcommand{\pcm}{\mbox{ ${\rm cm}^{-3}$}}

\normalsize

\begin{document}
\title{The X-ray Remnant of SN1987A}


\author{
David N. Burrows\altaffilmark{1},
Eli Michael\altaffilmark{2},
Una Hwang\altaffilmark{3},
Richard McCray\altaffilmark{2},
Roger A. Chevalier\altaffilmark{4},
Robert Petre\altaffilmark{3},
Gordon P. Garmire\altaffilmark{1},
Stephen S. Holt\altaffilmark{3},
John A. Nousek\altaffilmark{1}}

\altaffiltext{1}{Department of Astronomy \& Astrophysics,
	Penn State University, University Park, PA 16802;
	burrows@astro.psu.edu}
\altaffiltext{2}{JILA, University of Colorado, Boulder, CO 80309-0440}
\altaffiltext{3}{NASA/Goddard Space Flight Center, Greenbelt, MD 20771}
\altaffiltext{4}{Dept. of Astronomy, University of Virginia,
	P.O. Box 3818, Charlottesville, VA 22903}

\begin{abstract}
We present high resolution \Chandra\ observations of the remnant of
SN1987A in the Large Magellanic Cloud. The high angular resolution of
the \Chandra\ {\it X-ray Observatory (CXO)} permits us to resolve the
X-ray remnant. We find that the remnant is shell-like in morphology,
with X-ray peaks associated with some of the optical hot spots seen
in {\it HST} images.
The X-ray light curve has departed from the linear flux
increase observed by ROSAT, with a 0.5-2.0 keV luminosity of
$1.5 \times 10^{35}$ erg/s
in January 2000.
We set an upper limit of $2.3 \times 10^{34}$ ergs/s on the luminosity
of any embedded central source (0.5 - 2 keV).
We also present a high resolution spectrum, showing that the X-ray
emission is thermal in origin and is dominated by highly ionized species
of O, Ne, Mg, and Si.
\end{abstract}


\section{INTRODUCTION}
Supernova 1987A (SN1987A) in the Large Magellanic Cloud was the first
supernova explosion visible to the naked eye in nearly 400 years.
Because it is located in the nearest galaxy, it provides an unprecedented
opportunity to study the early evolution of a supernova remnant (SNR)
at a known distance (50 kpc; Andreani \etal\ 1987) with modern ground- and
space-based astronomical instruments.
As a result, it has been studied across the entire electromagnetic spectrum,
and is the only SN explosion from which a neutrino burst has been detected
(Koshiba \etal\ 1987), confirming our theoretical models of stellar collapse.
SN1987A is also the only remnant for which the stellar type of the progenitor star
is known from pre-explosion observations: the progenitor was the blue
supergiant Sanduleak $-69\deg 202$, a type B3 I star
(Kirshner \etal\ 1987; Sonneborn, Altner, \& Kirshner 1987).

The optical remnant has been studied intensively, both with ground-based
instruments and with the HST. HST images show a bright
elliptical ring ($1.7 \times 1.2$ arcseconds) surrounding the optical
remnant. This inner ring has been interpreted as a ring of high-density
material in the equatorial plane of the progenitor, caused by the impact
of the high velocity stellar wind of the blue supergiant progenitor with its
earlier, slower red giant wind, and ionized by the UV flash from the
supernova explosion 13 years ago (Burrows \etal\ 1995).

In recent years, HST observations of SN1987A have revealed a gradually
brightening ``hot spot'' on the inner ring at a position angle of 29
degrees (Garnavich \etal\ 2000).
This hot spot has been interpreted as evidence of shock heating
at the point of first contact of the supernova blast wave with
the relatively dense ($n \sim 10^4 \pcm$) inner ring
(Michael \etal\ 2000).
When the shock front engulfs the remainder of the inner ring, the
remnant is expected to brighten by as much as three orders of magnitude
at optical, UV, and X-ray wavelengths (Luo, McCray, \& Slavin 1994;
Borkowski, Blondin, \& McCray 1997b).
In the past several months, HST observations have shown that at least
four new optical hot spots have appeared along the inner eastern edge
of the inner ring (Bouchet \etal\ 2000; Lawrence \etal\ 2000;
Maran, Pun, \& Sonneborn 2000;
Garnavich, Kirshner, \& Challis 2000), providing further evidence that
this event is well underway.
A recent HST image of SN1987A is shown in Figure~\ref{fig:SN1987A}a.

At radio wavelengths, ATCA observations have shown a shell-like remnant
(Gaensler \etal\ 1997) that first appeared about 1300 days after explosion
and has been brightening and expanding steadily ever since
(Manchester \etal\ 2000).
An image made on 9 September 1999 is shown in Figure~\ref{fig:SN1987A}b.
The remnant has an irregular ring structure peaking just inside the
optical inner ring, with an eastern lobe about twice as bright as the
western lobe.
The radio emission fills the inner ring.

SN1987A was first detected in the soft X-ray band in Feb 1991
(1448 days after the explosion) by the {\it ROSAT} satellite, and has been
brightening steadily since then (Hasinger, Aschenbach, \& Tr\"{u}mper 1996).
With only 318 photons and the limited energy resolution of the
{\it ROSAT} PSPC, Hasinger \etal\ were unable to distinguish between simple
continuum models and more complex thermal plasma models of the spectrum.

We present the first \chandra\ images and spectra of SN1987A.  Although the
statistics are limited for these observations, we have resolved the remnant
in X-rays and have obtained a high resolution spectrum that is dominated by
thermal emission from highly ionized lines characteristic of temperatures
above $10^7$~K.  The luminosity has increased by a factor of three
since 1995, but there is no evidence yet for an embedded point source.

\section{IMAGE ANALYSIS}

We have observed SN1987A twice with the {\it Chandra X-ray Observatory}
(Weisskopf, O'Dell, \& van Speybroeck 1996) as part of the {\it Chandra}
Guaranteed Time Observations (GTO) program (Figure~\ref{fig:SN1987A}c
and \ref{fig:SN1987A}d). {\it Chandra} has unprecedented angular resolution
(0.5 arcseconds HPD) in the X-ray band.
It has two focal plane cameras and two sets of transmission gratings that
can be inserted into the optical path.
Our observations utilized the Advanced CCD Imaging Spectrometer
(ACIS; Garmire \etal\ 2000), which takes high-resolution images with
moderate non-dispersive energy resolution, and which can also serve as a
readout device for the transmission gratings.
Our first observation was made on 6~October~1999 using the High Energy
Transmission Grating and the ACIS-S detector array.
The 120 ks observation produced an undispersed zero-order image of the
remnant with moderate energy resolution, and a set of dispersed spectra with
high energy resolution.
The second observation, made on 17 January 2000, used the ACIS S3 detector,
and provided 10.7 ks of data with moderate nondispersive energy resolution.

\setcounter{footnote}{0}
The data were processed and filtered to reduce the background.
We restricted photon energies to 0.3 - 8 keV and used ASCA grade 02346 events.
In order to obtain the highest possible spatial resolution, we turned off
the pixel randomization introduced by the standard data processing
system\footnote{The pipeline software adds a random
offset of between
$\pm \frac{1}{2}$ pixel to
the sky position of each photon to prevent the appearance of
Moir\'{e} patterns in \chandra\ images.
This is unnecessary for long \chandra\
observations, because the spacecraft dither accomplishes the same thing.}.
The light curves were examined for times of high background that could
contaminate our results, but none were found.
The first observation contained 696 photons meeting our criteria in the
zero-order image, while the second observation had 583 photons.
The astrometry of the raw images was adjusted by hand to agree with the
reference system of Reynolds et al. (1995), who determined a position of
(05:35:27.97, -69:16:11.09) for SN1987A based on Hipparcos and VLBI data.

The optical and radio remnants measure roughly $1.7'' \times 1.2''$ on the
sky, as do the raw X-ray images.
This is barely resolved by the {\it Chandra}/ACIS instrument,
which samples the $\sim 0.5''$ telescope image with $0.492''$ detector
pixels.
Because {\it Chandra} moves slowly across the sky during each observation,
the placement of ACIS detector pixels on the sky is not fixed during an
observation, allowing us to deconvolve our data to obtain higher resolution.
The ACIS data were binned into 0.125 arcsecond sky pixels (with
the center of the CCD pixel projected into the corresponding sky pixel),
and the resulting images were then deconvolved with a Maximum Likelihood
algorithm (20 iterations; Richardson 1972; Lucy 1974), using an on-axis
point spread function obtained with the MARX ray tracing software package.
Smoothed versions of the deconvolved images are shown in
Figure~\ref{fig:SN1987A}c and \ref{fig:SN1987A}d, and have an effective
resolution of about 0.3 arcseconds (FWHM).
The X-ray remnant size, measured between peaks of the shell, is
$1.2 \times 1.0$ arcseconds.

The SNR is generally shell-like, with the X-ray emission peaking just
inside the optical inner ring, as expected on the basis of
current models (Chevalier \&
Dwarkadas 1995; Borkowski, Blondin, \& McCray 1997a).
The X-ray emission is brightest in the eastern quadrant, where the new
hot spots are appearing.
We believe that the observed X-ray emission comes primarily from the
shocked supernova debris and shocked circumstellar gas in a shell between
the reverse shock and the blast wave (Borkowski, Blondin, \& McCray 1997a).
The blast wave is propagating with velocity $\sim 4,000$ km/s into a
relatively low density ($n \sim 10^2 \pcm$) HII region.
The presence of the reverse shock is indicated by spectroscopic observations
made with HST/STIS, which show high velocity Ly$\alpha$ and H$\alpha$
emission coming from a surface interior to the circumstellar ring where
the freely expanding supernova debris is suddenly slowed from
$\sim 15,000$ km/s to $\sim 3,000$ km/s (Michael \etal\ 1998).
The nonthermal radio emission probably comes from relativistic particles
in the same zone.

The X-ray images in Figure~\ref{fig:SN1987A} are overlaid with contours
of the HST H$\alpha$ emission.
Some of the X-ray hot spots agree well with optical hot spot positions
(small differences of order 0.1 arcseconds
are not significant);
others do not.
Differences between the two X-ray observations are
probably dominated by statistical variations;  a $1\sigma$ variation in
these images corresponds to about 50 counts, or about 8\% of the total,
whereas the details of the deconvolved images typically involve fewer
counts than that.
Longer observations planned for Cycle 2 will provide images with greatly
improved statistics.

Are we observing X-rays from the optical hot spots themselves?
The evidence on this question is tantalizing but ambiguous.
The X-ray images exhibit peaks
in the general vicinity of the optical hot spots, but the positional
agreement
is not perfect; further, some of this structure is similar to that seen
in the ATCA images, which does not seem to be related to the optical hot
spots.
STIS spectra of Spot 1
show line broadening consistent with shock velocities in the range
100 - 300 km/s (Michael \etal\ 2000).
Shocks with such velocities produce X-rays with spectra much softer than
the observed spectrum, and shock models with parameters chosen to fit
observations of UV emission lines from Spot 1 predict X-ray fluxes only
$\sim 0.5 \%$ of the total X-ray flux observed by Chandra.
On the other hand, a greater flux of X-rays may be emitted by faster
($> 300$ km/s) shocks that are also propagating into the hot spots.
The X-ray emitting gas behind such shocks may not have had time to cool
by radiation and emit observable optical and UV emission lines.

We see no evidence for a point source of X-rays in the interior of SN1987A.
We have performed a Monte Carlo simulation to estimate our point source
sensitivity.
In this simulation, we added a point source (convolved with the PSF,
with photon statistics) to the observed photon image in eighth-arcsecond
pixels,
rebinned to half-arcsecond pixels, and compared the result to the original image
using a $\chi^2$ test.
The 90\% confidence limit on a central point source is
15\% of the total SNR flux.
This allows us to place an upper limit of $2.3 \times 10^{34}$ ergs/s
on the luminosity of any central point source (0.5 - 2 keV).
This limit is not surprising in view of calculations showing that the
debris should still be opaque to soft X-rays
(Fransson \& Chevalier 1987).
(By comparison, Suntzeff \etal\ 1992 set an upper limit of
$8 \times 10^{36}$ erg/s on the contribution of a compact central source
to the optical light curve on day 1500.)

\section{LIGHT CURVE}

The long-term light curve of SN1987A in soft X-rays is shown in
Figure~\ref{fig:light_curve}, where we have plotted observed fluxes
(0.5 - 2.0 keV) calculated
from our new observations together with earlier ROSAT measurements
(Hasinger, Aschenbach, \& Tr\"{u}mper 1996).
ACIS count rates were converted to observed fluxes
and source luminosities by fitting the undispersed
spectra to plane parallel shock models.
The ROSAT count rates were converted to observed fluxes using the same models
that fit the ACIS data.
The X-ray flux continues to increase monotonically, and is
now substantially higher than the linear trend of the ROSAT data.
This may be due to the interaction of the blast wave with parts of the
equatorial ring, or may simply reflect higher densities in the regions now
being shocked.
The fact that both the X-ray and radio light curves
continue to increase monotonically,
combined with the morphological similarity of the X-ray and radio images,
supports the notion that the shocks responsible for the X-ray emission
are also responsible for accelerating the relativistic electrons
that emit the radio waves 
(Chevalier \& Dwarkadas 1995).
The X-ray luminosity in Jan 2000 was $1.5 \times 10^{35}$ ergs/s (0.5-2.0 keV), or
$1.9 \times 10^{35}$ ergs/s (0.5-10 keV), for a distance of 50 kpc.

\section{DISPERSED SPECTRUM}

Our first observation produced a dispersed high resolution spectrum,
which is shown in Figure~\ref{fig:dispersed_spectrum}.
X-ray events with inferred energy corresponding to their position along the
dispersed spectrum were extracted using a 3 arcsecond wide source region,
with 7.5 arcsecond background regions on either side,
to optimize the signal to noise ratio.
Photons from the positive and negative first orders
of both the MEG and HEG were combined to
produce the spectrum shown in Figure~\ref{fig:dispersed_spectrum}.

The spectrum is dominated by bright lines from the Helium-like and
Hydrogen-like ionization stages of O, Ne, Mg, and Si. Preliminary fits
to plane parallel shock models with constant electron temperature
(vpshock in XSPEC V11) and abundances appropriate for the inner
circumstellar ring (Lundqvist \& Fransson 1996) and the LMC (Russell
\& Dopita 1992; Hughes, Hayashi, \& Koyama 1998) give electron
temperatures of $kT \approx 3$ keV and ionization timescales of $nt \approx 7
\times 10^{10}$ cm$^{-3}$ s.  This electron temperature is much lower
than the post-shock proton temperature for a $4000$ km s$^{-1}$ shock
($T_p \approx 30$ keV) because the time for temperature equilibration
through Coulomb interactions is of order $5000$ years.
The shock
ionization timescale, $nt$, is just about what would be expected for
a blast wave
that has propagated for $t \sim 6$ years into a gas having pre-shock
density $n \sim 10^2 \pcm$.

The X-ray line shapes are determined by a combination of thermal and bulk
motion of the rapidly expanding young remnant, and can be used to constrain
the shock velocity and possibly its geometry.
We have co-added four of the brightest
isolated X-ray lines to examine the line profile.
Preliminary analysis of this profile appears to be consistent with an
equatorially brightened shell expanding with velocity $\sim 4000$~km/s,
plus emission
from lower velocity shocks entering the hotspots.
We defer detailed discussion to a future paper on spectroscopic analysis of
these data.


\section{CONCLUSIONS}

The {\it Chandra X-ray Observatory} has opened up an exciting new era of
high spatial and spectral resolution X-ray astronomy.
Our preliminary analysis demonstrates that the X-ray emission from SN1987A is beginning to
show the effects of the interaction of the blast wave with the circumstellar
medium, signalling the birth of a supernova remnant.
For the first time, we can follow the development of a young
supernova remnant (SNR) and confront theoretical models of SNR evolution
with detailed X-ray observations.

SN1987A is a rapidly evolving object:
the flux has increased by a factor of 3 in the past five years, and is
expected to increase more dramatically in the near future.
We predict that emission from
hot spots, which now accounts for perhaps 20\%	of the X-ray flux, will
dominate the X-ray flux within 2-3 years.
The rapid timescale of changes requires periodic monitoring of SN1987A,
using Chandra/ACIS to produce images of the remnant, XMM-Newton to
produce high-resolution spectra, and HST to monitor the optical and UV images
and spectra of the hot spots.

\acknowledgements
This work was supported by NASA contract NAS8-38252 and NASA grant NGT5-80.
Support was also provided by NASA through grant GO-08243 from the Space
Telescope Science Institute, which is operated by the Association of
Universities for Research in Astronomy, Inc., under NASA contract NAS5-26555.
We are grateful to Bryan Gaensler and to Peter Challis and the SINS team for
providing their data for comparison with our {\it Chandra} observations,
and to the HETG team at MIT for assistance with the analysis of the
dispersed spectrum.
The superb performance of the \Chandra\ observatory is the result of
years of work by thousands of scientists, engineers, and managers, and
we are deeply indebted to their efforts.
Finally, we acknowledge the very useful comments of our referee, which resulted
in significant improvements to this paper.


\clearpage

{\bf References}
\begin{verse}

Andreani, P., Ferlet, R., \& Vidal-Madjar, A. 1987, Nature, 326, 770

Borkowski, K. J., Blondin, J. M., \& McCray, R. 1997a, \apj, 476, L31

Borkowski, K. J., Blondin, J. M., \& McCray, R. 1997b, \apj, 477, 281

Bouchet, P. \etal\ 2000, \iauc, 7354

Burrows, C. J., \etal\ 1995, \apj, 452, 680

Chevalier, R. A., \& Dwarkadas, V. V. 1995, \apj, 452, L45

Fransson, C., \& Chevalier, R. A. 1987, \apj, 322, L15

Gaensler, B., \etal\ 1997, \apj, 479, 845

Garmire, G. P., \etal\ 2000, \apjs, in preparation

Garnavich, P., Kirshner, R., \& Challis, P. 2000, \iauc, 7360

Garnavich, P., \etal\ 2000, \apj, submitted

Hasinger, G., Aschenbach, B., \& Tr\"{u}mper, J. 1996, \aa, 312, L9

Hughes, J. P., Hayashi, I., \& Koyama, K. 1998, \apj, 505, 732

Kirshner, R. P., \etal\ 1987, \apj, 320, 602

Koshiba, M., \etal\ 1987, \iauc, 4338

Lawrence, S., Sugarman, B., Bouchet, P., Crotts, A., Ugleshich, R.,
\& Heathcote, S. 2000, \apj, in press

Lucy, L. B. 1974, \aj, 79, 745

Lundqvist, P. \& Fransson, C. 1996, \apj, 464, 924

Luo, D., McCray, R., \& Slavin, J. 1994, \apj, 430, 264

Manchester, R., \etal\ 2000, PASA, in press

Maran, S., Pun, C. S. J., \& Sonneborn, G. 2000, \iauc, 7359

Michael, E., \etal\ 1998, \apj, 509, L117

Michael, E., \etal\ 2000, \apj, submitted

Reynolds, J. E., \etal\ 1995, \aa, 304, 116

Richardson, W. H. 1972, {\it J. Opt. Soc. Am.}, 62, 55

 Russell, S. C., \& Dopita, M. A. 1992, \apj, 384, 508

Sonneborn, G., Altner, B., \& Kirshner, R. P. 1987, \apj, 323, L35

Suntzeff, N. B., \etal\ 1992, \apj, 384, L33

Weisskopf, M. C., O'Dell, S. L., \& van Speybroeck, L. P. 1996, Proc. SPIE, 2805, 2

\end{verse}

\newpage



\begin{figure}[h]
\caption{Four recent images of SN1987A: a: HST image (log scaling)
taken on
2 February 2000
(P. Garnavich, R. Kirshner, \& P. Challis and the SINS team);
b: Deconvolved ATCA 8 GHz image (square-root scaling) from 9 September 1999
(Manchester \etal\ 2000);
c: Deconvolved ACIS image (linear scaling) from 6 October;
d: Deconvolved ACIS image (linear scaling) from 17 January 2000.
The images were made with identical pixel size and orientation (north is up).
Contour overlays in panels b, c, and d are the optical ring from panel a.
The ATCA data were kindly made available to us for this comparison by
Bryan Gaensler and Dick Manchester.
The Australia Telescope Compact Array is funded by the
Commonwealth of Australia for operation as a National Facility managed
by CSIRO.}
\label{fig:SN1987A}
\end{figure}


\begin{figure}[h]
\caption{SN1987A light curves in the 4.7 GHz (top) and soft X-ray (bottom) bands.  The X-ray light
curves are for the observed fluxes in the 0.5 - 2.0 keV band.  The
solid lines are the best-fit linear trends, fitted to the entire ATCA and ROSAT
data sets.}
\label{fig:light_curve}
\end{figure}

\begin{figure}[h]
\caption{Dispersed X-ray spectrum of SN1987A from the $\pm 1^{\rm st}$ orders of
the MEG and HEG spectra. The spectrum has been smoothed with a Gaussian having FWHM $=$ 0.03 keV.}
\label{fig:dispersed_spectrum}
\end{figure}


\end{document}